\documentclass[12pt]{article}
\usepackage{euscript,amsfonts,amssymb,amsmath,color}
\usepackage{graphicx}
\usepackage{caption}
\usepackage{subcaption}
\usepackage{soul}
\usepackage{eurosym}

\newcommand{\mathstack}[1]{\mathop{#1}\limits}
\newcommand{\rarely}[1]{}
\def\tag(#1,#2)[#3]#4{\put(#1,#2){\makebox(0,0)[#3]{#4}}}
\newcommand{\converges}[2]{{
     \setbox0\hbox{${\scriptscriptstyle #1\ }$}
     \setbox2\hbox{${\scriptscriptstyle #2\ }$}
     \setbox4\hbox{$\rightarrow$}
     \ifdim\wd0>\wd2
          \ifdim\wd0>\wd4
          \,\mathstack{\hbox to
\wd0{\rightarrowfill}}_{\scriptscriptstyle#1}^{\scriptscriptstyle #2}\,
            \else
\newcommand\zfrac[2]{{\textstyle\frac{#1}{#2}}}

\mathstack{\rightarrow}_{\scriptscriptstyle#1}^{\scriptscriptstyl e#2}\,
            \fi
       \else
            \ifdim\wd2>\wd4
            \,\mathstack{\hbox to
\wd2{\rightarrowfill}}_{\scriptscriptstyle#1}^{\scriptscriptstyle #2}\,
            \else\

\mathstack{\rightarrow}_{\scriptscriptstyle#1}^{\scriptscriptstyl e#2}\,
            \fi
       \fi}}

\newcommand{\Prob}{{\rm P}}

\newtheorem{definition}{Definition}

\newtheorem{corollary}{Corollary}

\newtheorem{proposition}{Proposition}

\begin{document}

\begin{center}
{\Large \bf Continuity and Monotonicity of Preferences and Probabilistic Equivalence}

by

Sushil Bikhchandani\footnote{Anderson School of Management, UCLA, sbikhcha@anderson.ucla.edu} and Uzi Segal\footnote{Dept.\ of Economics, Boston College, segalu@bc.edu}

\today

\end{center}

\bigskip

Let ${\cal L}$ be the set of real finite-valued random variables over $(S,\Sigma,\Prob)$ with $S=[0,1]$, $\Sigma$ being the standard Borel $\sigma$ algebra on $S$, $\Prob = \mu$, the Lebesgue measure, and the set of outcomes being the bounded interval $[\underline x, \bar x]$. The decision maker has a preference relation $\succeq$ over $\cal L$. In the sequel, we denote events by $S_i$ and $T_i$.

\begin{definition}
\label{d:1}
The continuous function $\psi:[\underline x, \bar x]\times [\underline x, \bar x] \to \Re$ is a {\em regret function} if for all $x$, $\psi(x,x)=0$, $\psi(x,y)$ is strictly increasing in $x$, and strictly decreasing in $y$.
\end{definition}

If in some event $X$ yields $x$ and $Y$ yields $y$ then $\psi(x,y)$ is a measure of the decision maker's {\it ex post} feelings (of regret if $x<y$ or rejoicing if $x>y$) about the choice of $X$ over $Y$. This leads to the next definition:

\begin{definition}
\label{d:2}
Let $X,Y\in\cal L$ where $X = (x_1,S_1;\ldots;x_n,S_n)$ and $Y=(y_1,S_1;$ $\ldots;y_n,S_n)$. The {\em regret lottery} evaluating the choice of $X$ over $Y$ is $$\Psi(X,Y) =  (\psi(x_1,y_1),p_1;\ldots;\psi(x_n,y_n),p_n)$$
where $p_i = \Prob(S_i)$, $i=1,\ldots,n$. Denote the set of regret lotteries by ${\cal R} = \{\Psi(X,Y):X,Y \in {\cal L}\}$.
\end{definition}

\begin{definition}
\label{d:3}
The preference relation $\succeq$ is {\em regret based} if there is a regret function $\psi$ and a continuous functional $V$ which is defined over regret lotteries such that for any $X, Y\in \cal L$
\[
X\succeq Y \quad \mbox{if and only if } \quad V(\Psi(X,Y)) \geqslant 0
\]
\end{definition}

\rarely{

Under the assumption of continuity of $\succeq$ with respect to the weak topology (or convergence in distribution), Proposition 1 is superfluous. The aim of this note is to provide a proof of Proposition 1 under the weaker assumption of continuity of $\succeq$ with respect to convergence of random variables in probability. Although monotonicity of $\succeq$ is not directly used in the proof of Proposition 1 below, state-wise monotonicity of $\succeq$ is assumed in the paper -- it is implicit in the monotonicity assumption on $\psi$, the regret function. 

We then show in a corollary below that Proposition~1 implies a strengthening these assumptions to continuity with respect to convergence in distribution and monotonicity with respect to first-order stochastic dominance. (These stronger assumptions of continuity and monotonicity are used in ``Transitive Regret.'')

Let ${\cal L}$ be the set of real finite-valued random variables over $(S,\Sigma,\Prob)$ with $S=[0,1]$, $\Sigma$ being the standard Borel $\sigma$ algebra on $S$, $\Prob = \mu$, the Lebesgue measure, and the set of outcomes being the bounded interval $[\underline x, \bar x]$. The decision maker has a preference relation $\succeq$ over $\cal L$. In the sequel, we denote events by $S_i$ and $T_i$.

}

Let $X = (x_1,S_1;\ldots;x_n,S_n)\in \cal L$, and let  $X^k = (x^k_1,S^k_1;\ldots;x^k_m,S^k_m)\in~{\cal L}$ be a sequence of random variables. %\red{Let $F$ and $F^k$ be the cdfs of $X$ and $X^k$.} 
The sequence $X^k$ {\sl convergences in probability} to $X$, denoted $X^k\stackrel{p}{\longrightarrow} X$, if $\forall \varepsilon>0$,
\begin{align*}
\lim_{k\to\infty}\Prob\big(\big|X^k-X\big|\geqslant\varepsilon\big) = 0
\end{align*}
(See Billingsley~\cite[p.\ 274]{Bil79}.)

A preference relation $\succeq$ is {\sl continuous w.r.t.\ convergence in probability} if $X^k\succeq Y$ for all $k$ and $X^k\stackrel{p}{\longrightarrow} X$ implies $X\succeq Y$ and $Y\succeq X^k$ for all $k$ and $X^k\stackrel{p}{\longrightarrow} X$ implies $Y\succeq X$.

A preference relation $\succeq$ satisfies {\sl state-wise monotonicity} if for any $X = (x_1,S_1;\ldots;x_n,S_n)$ and $Y=(y_1,S_1;…;y_n,S_n)$ where for all $i$, $ x_i\geqslant y_i$ with at least one strict inequality then $X\succ Y$.

As pointed out by Chang and Liu~\cite{ChaLiu24}, Proposition 1 in~\cite{BikSeg11} is unclear. This proposition is the first step in proving the main result, Theorem 1. So it is implicit that the assumptions of Theorem 1 are invoked in proving Proposition 1, but we did not define the notion of continuity and monotonicity that the preference relation satisfies. Implicitly we assumed continuity w.r.t. convergence in distribution, which makes the proposition trivial. Here we show that it holds even if continuity wrt convergence in probability is assumed. % \red{Throughout we use the notation in \cite{BikSeg11}. }

\begin{proposition} {\rm (Probabilistic equivalence)}. Let $\succeq$ be a complete, transitive, continuous w.r.t.\ convergence in probability, and state-wise  monotonic, regret-based preference relation over $\cal L$. For any two random variables $X,Y\in{\cal L}$, if $F_X=F_Y$, then $X\sim Y$.
\end{proposition}

\noindent
{\bf Proof:} Let $X= (x_1,S_1;\ldots;x_n,S_n)$ and $Y = (y_1,S_1';\ldots;$ $y_n,S_n')$ be such that $F_X=F_Y$.
\medskip

\noindent
{\sf Case 1:} $S_i=S_i'$ and $\Prob(S_i) = \frac1n$, $i=1,\ldots,n$. Then there is a permutation $\hat\pi$ such that $Y = \hat\pi(X)$. Obviously, $\Psi(X,\hat\pi(X)) = \Psi(\hat\pi^i(X),$ $\hat\pi^{i+1}(X))$.\footnote{This is the only place where the assumption of regret-based $\succeq$ is used in the proof. Thus, the proposition can be proved under a weaker assumption that $X\succ \pi(X)$ implies $\pi(X) \succ \pi^2 (X)$.} Hence, as there exists $m \leqslant n!$ such that $\hat\pi^m(X) = X$, it follows by transitivity that for all $i$, $X \sim \hat\pi^i(X)$. In particular, $X \sim Y$.
\medskip

\noindent
{\sf Case 2:} For all $i,j$, $\Prob(S_i \cap S'_j)$ is a rational number. Let $N$ be a common denominator of all these fractions. $X$ and $Y$ can now be written as in case~1 with equiprobable events $T_1,\ldots,T_N$.
\medskip

\medskip
\noindent
{\sf Case 3:} There exist $i$ and $j$, such that $P(S_i\cap S'_j)$ is irrational. For $x_1<\ldots<x_n$ and $y_1<\ldots<y_n$, let $X=(x_1,S_1;\ldots;x_n,S_n)$ and $Y=(y_1,S'_1;$ $\ldots;y_n,S'_n)$ be such that $F_X=F_Y$. Then $x_i=y_i$ and $p_i:=P(S_i)=P(S'_i)$, $i=1,\ldots,n$. Let $T_1,\ldots,T_m$ be the set of intersections $\{S_i\cap S'_j:P(S_i\cap S'_j)>0\}$. Clearly, $\sum_j\{P(T_j):X(T_j)=x_i\}=\sum_j\{P(T_j):Y(T_j)=x_i\}=p_i$, $i=1,\ldots,n$.

For $k=1,2,\ldots$, define $\nu(T_j,k)$ such that
\begin{eqnarray*}
\frac{\nu(T_i,k)}{2^k}< P(T_j)\leqslant\frac{\nu(T_j,k)+1}{2^k}
\end{eqnarray*}
For $k$ such that $\frac1{2^k}<\min_j\{P(T_j)\}$, define a partition $T^k=\{T^k_{jh}:j=1,\ldots,m,\,h=1,\ldots,\nu(T_j,k)\}$ of $[0,1]$ satisfying
\begin{itemize}
\item $\sum_{h=1}^{\nu(T_j,k)}P(T^k_{jh})=P(T_j)$, $j=1,\ldots,m$.
\item For $j=1,\ldots,m$ and $h=1,\ldots,\nu(T_j,k)-1$, $P(T^k_{jh})=\frac1{2^k}$.
\end{itemize}
That is, $T^k$ partitions each $T_j$ into $\nu(T_j,k)-1$ events with probability $\frac1{2^k}$ each, and one event with probablity not greater than $\frac1{2^k}$. Define $X^k,Y^k$ such that
\begin{itemize}
\item For $j=1,\ldots,m$ and $h=1,\ldots,\nu(T_j,k)-1$, $X^k=X$ and $Y^k=Y$.
\item For $j=1,\ldots,m$ and $h=\nu(T_j,k)$, $X^k=Y^k=c$, where $c\not\in\{x_1,\ldots,x_n\}$.
\end{itemize}
Observe that $X^k$ disagrees with $X$ and $Y^k$ disagrees with $Y$ on at most $m$ elements of $T^k$. Hence, for every $i$, $P(X^k=x_i)\geqslant P(X=x_i)-\frac m{2^k}$ and $P(Y^k=x_i)\geqslant P(Y=x_i)-\frac m{2^k}$. Note that by definition, $P(X^k=x_i)\leqslant P(X=x_i)$ and $P(Y^k=x_i)\leqslant P(Y=x_i)$. It thus follows that 
\begin{align*}
|P(X^k=x_i)-P(Y^k=x_i)|\leqslant\frac m{2^k}, \qquad\forall i
\end{align*}
(Recall that $P(X=x_i)=P(Y=x_i)$.) Modify $X^k$ and $Y^k$ as follows. If $d=P(X^k=x_i)-P(Y^k=x_i)>0$, then change $d2^k$ elements of the partition $T^k$ where $X^k$ yields $x_i$ to yield $c$ instead, and if $d=P(Y^k=x_i)-P(X^k=x_i)>0$, then change $d2^k$ elements of $T^k$ where $Y^k$ yields $x_i$ to yield $c$ instead. Denote the new random variables $\bar X^k$ and $\bar Y^k$. Observe that:
\begin{description}
\item (a) $F_{\bar X^k}=F_{\bar Y^k}$
\item (b) $P(\bar X^k\neq X)\leqslant\frac{m^2}{2^k}$, $P(\bar Y^k\neq Y)\leqslant\frac{m^2}{2^k}$, hence $\bar X^k\stackrel{p}{\longrightarrow} X$ and $\bar Y^k\stackrel{p}{\longrightarrow} Y$
\item (c) For every $i$, $P(\bar X^k=x_i)=P(\bar Y^k=x_i)=\frac{\ell_i}{2^k}$ for some integer $\ell_i$, therefore $P(\bar X^k=c)=P(\bar Y^k=c)=\frac\ell{2^k}$ for some integer $\ell$.
\end{description}
By (a), (c), and case 2, $\bar X^k\sim\bar Y^k$, and by (b) and continuity, $X\sim Y$.~\hfill~$\blacksquare$

\bigskip

Proposition~1 implies that $\succeq$  satisfies a stronger form of continuity and monotonicity, as shown next.

Let $F_X$ be the cdf of $X\in \cal L$ and $F_{X^k}$ be the cdf of $X^k\in \cal L$. 
A sequence of random variables $X^k$ {\sl converges in distribution}  to $X$, denoted $X^k \stackrel{d}{\longrightarrow} X$, if 
\begin{align*}
\lim_{k\to\infty} F_{X^k}(x)= F_X(x) 
\end{align*}
at every $x$ at which $F_X$ is continuous (see Billingsley~\cite[p.\ 338]{Bil79}).

A preference relation $\succeq$ is {\sl continuous w.r.t.\ convergence in distribution} if $X^k\succeq Y$ for all $k$ and $X^k\stackrel{d}{\longrightarrow} X$ implies $X\succeq Y$ and $Y\succeq X^k$ for all $k$ and $X^k\stackrel{d}{\longrightarrow} X$ implies $Y\succeq X$.

%\bigskip
\noindent
\begin{corollary}  Let $\succeq$ be a regret-based preference relation over $\cal L$. Assume that $\succeq$ satisfies the assumptions of Proposition~1. Then $\succeq$ is (i) monotonic w.r.t.\ first-order stochastic dominance (FOSD) and (ii) continuous w.r.t.\ convergence in distribution.
\end{corollary}

\noindent
{\bf Proof:} (i) Let $X= (x_1,S_1;\ldots;x_n,S_n)$ and $Y = (y_1,S_1';\ldots;$ $y_n,S_n')$ be such that  $F_X$ strictly dominates $F_Y$ by FOSD. One can construct two random variables $X',\, Y'$ with cdfs $F_X$ and $F_Y$ respectively such that  $X' = (x_1',T_1;\ldots;x_m',T_m)$, $Y'=(y_1',T_1;…;y_m',T_m)$, $ x_i'\geqslant y_i'$ for all $i$. Observe that each $x_i'\in\{x_1,\ldots, x_n\}$ and each $y_i'\in\{y_1,\ldots, y_n\}$. As $F_X$ strictly dominates $F_Y$, for at least one $i$, we have $ x_i'> y_i'$.  Therefore, state-wise monotonicity implies that $X'\succ Y'$. By Proposition 1, $X'\sim X$, $Y'\sim Y$ and by transitivity, $X\succ Y$. Thus $\succeq$ satisfies monotonicity w.r.t.\ FOSD. 

\smallskip
\noindent
(ii) Suppose that $X^k\stackrel{d}{\longrightarrow} X$ and that $X^k \succeq Y$ for all $k$. We show that $X\succeq Y$. By Skohorod's Theorem (see Billingsley~\cite[p.\ 343]{Bil79}), there exists a sequence of random variables $\bar X^k$ such that $\bar X^k\stackrel{p}{\longrightarrow} X$ and $F_{\bar X^k}=F_{ X^k}$. By Proposition 1, $\bar X^k\sim X^k$. Therefore, $X^k \succeq Y$ for all $k$ and transitivity imply that $\bar X^k \succeq Y$ for all $k$. Continuity w.r.t.\ convergence in probability implies that $X\succeq Y$. 
\hfill$\blacksquare$

\end{document}